\documentclass[preprint,11pt,number]{elsarticle}

\usepackage[T1]{fontenc}
\usepackage{lmodern}
\usepackage[a4paper,textwidth=460pt,top=22mm,bottom=22mm]{geometry}
\usepackage{amsmath,amssymb}
\usepackage{array,booktabs,tabularx}
\usepackage[colorlinks=true,linkcolor=blue,citecolor=blue,urlcolor=blue]{hyperref}
\usepackage{doi}
\makeatletter
\def\ps@pprintTitle{%
  \let\@oddhead\@empty
  \let\@evenhead\@empty
  \def\@oddfoot{\reset@font\hfil\thepage\hfil}%
  \let\@evenfoot\@oddfoot}
\makeatother

\biboptions{sort&compress}

\hypersetup{
  pdftitle={The Zero Mode Constraint in the Dirac-Bergmann Analysis of Kadomtsev-Petviashvili Type Equations},
  pdfauthor={Ali Pazarci}
}

\newcommand{\pd}{\partial}
\newcommand{\dpx}{\partial_x^{-1}}
\newcommand{\vx}{\mathbf{x}}
\newcommand{\D}{\delta^{(2)}(\vx-\vx')}
\newcommand{\ct}{\tilde{c}_1}
\newcommand{\lt}{\tilde{\lambda}_1}
\newcommand{\dph}{\hat{\partial}_x^{-1}}
\newcommand{\Dx}{D_x^{-1}}
\newcommand{\av}[1]{\langle #1\rangle}

\begin{document}
\begin{frontmatter}

\title{The Zero Mode Constraint in the Dirac-Bergmann Analysis of Kadomtsev-Petviashvili Type Equations}

\author[physics,feza]{Ali Pazarci\corref{corresponding}}
\ead{ali.pazarci@std.bogazici.edu.tr}
\cortext[corresponding]{Corresponding author}
\address[physics]{Department of Physics, Bo\u{g}azi\c{c}i University, 34342 Bebek, Istanbul, T\"urkiye}
\address[feza]{Feza G\"ursey Center for Physics and Mathematics, Bo\u{g}azi\c{c}i University, 34684 \c{C}engelk\"oy, Istanbul, T\"urkiye}

\begin{abstract}
We derive the Hamiltonian formulation of the Kadomtsev-Petviashvili (KP) equation from a singular Lagrangian by means of the Dirac-Bergmann algorithm. The Lagrangian contains an auxiliary field, which leads to an irregular secondary constraint. On the branch that reproduces the KP equation, all constraints are second class, and the Dirac brackets give the known Hamiltonian and Poisson operator. We confirm these results by the Faddeev-Jackiw method. Then, we consider fields periodic in the first spatial coordinate. In this case, the zero modes of the derivative operator lead to an additional secondary constraint on the mean of the potential. This constraint replaces the inverse derivative by its zero mean counterpart, and the resulting Hamiltonian is in accordance with the one of Reiman and Semenov-Tyan-Shanskii, up to normalization conventions. For decaying fields, the same condition is necessary for rapidly decreasing potentials to remain rapidly decreasing. We show that this constraint appears for a general class of KP-type equations. In addition, we classify the smooth solutions of the auxiliary-field action. We also relate our results to an existing quantization of the KP equation, for which the potential Lagrangian provides an action.
\end{abstract}

\end{frontmatter}
%=====================================================================

\section{Introduction}

The Kadomtsev-Petviashvili (KP) equation was introduced to describe the stability of solitary waves in weakly dispersive media with weak transverse perturbations \cite{KP1970}. It is the natural (2+1)-dimensional generalization of the Korteweg-de Vries (KdV) equation, to which it reduces when the dependence on the transverse coordinate is neglected. The KP equation is completely integrable. It admits a Lax representation \cite{Dryuma1974,ZakharovShabat1974}, can be solved by the inverse scattering method, and possesses infinitely many conservation laws \cite{NMPZ1984,AblowitzClarkson1991}. Moreover, solutions of the KP equation can be obtained from the solutions of a singular manifold equation, which arises from a truncated Painlev\'e expansion \cite{Kudryashov1993}. Depending on the sign of the transverse dispersion term, one distinguishes two versions of the equation, KP-I and KP-II.

The Hamiltonian structure of the KdV equation was established by Gardner \cite{Gardner1971} and by Zakharov and Faddeev \cite{ZakharovFaddeev1971}, with the Poisson operator $\pd_x$. The Hamiltonian structure of the KP equation and of equations of KP type has also been studied \cite{NMPZ1984,ReymanSemenov,WuZhou2016}. Reiman and Semenov-Tyan-Shanskii obtained it from a Lie-Poisson bracket \cite{ReymanSemenov}. They also pointed out that a consistent Hamiltonian treatment of the KP equation requires a careful choice of the functional class. Bi-Hamiltonian structures were found for the dispersionless Boussinesq and Benney-Lax equations, which are reductions of the dispersionless KP equation \cite{GumralNutku1994}. Lipovskii used Dirac's theory of constrained systems to handle the relations among the scattering data of the KP-II equation with decreasing Cauchy data \cite{Lipovskii1986}. The Hamiltonian form of the KP-II equation was also used to study the transverse stability of its periodic waves \cite{HaragusLiPelinovsky2017}.

In this paper, we derive the Hamiltonian formulation of the KP equation from a Lagrangian density. The Lagrangian used here is linear in the velocities, as is usual for evolution equations of first order in time. Hence, its Hessian matrix vanishes, and the canonical momenta cannot be solved for the velocities. Therefore, passing to the Hamiltonian picture requires the theory of constrained systems developed by Dirac and Bergmann \cite{Dirac1950,AndersonBergmann1951,Dirac1964,HenneauxTeitelboim1992}. This theory has also been applied to second-order degenerate Lagrangians \cite{CagatayEsenGumral2018}. We also cross-check our results by the symplectic method of Faddeev and Jackiw \cite{FaddeevJackiw1988} in \ref{app:FJ}.

Dirac's theory was applied to the KdV equation by Nutku \cite{Nutku1984}. He later related the resulting symplectic form to the covariant Witten-Zuckerman symplectic current \cite{Nutku2000}. Olver showed that, for a constant-coefficient Hamiltonian operator, the canonical system obtained by Dirac's theory can be constructed directly \cite{Olver1986}. Nutku and Pavlov constructed multiple Lagrangians for integrable systems from their multi-Hamiltonian structures \cite{NutkuPavlov2002}. This approach was also used for integrable systems in more than two dimensions, such as the real Monge-Amp\`ere equations \cite{Nutku1996} and Pleba\'nski's second heavenly equation \cite{NeyziNutkuSheftel2005}. Since the Poisson operator of the KP equation is a constant multiple of $\pd_x$, the bracket obtained below is in accordance with Olver's construction. Dirac's approach was also used for the CH-KP equation \cite{CamassaFalquiSforza2026} and for coupled and octonionic KdV systems \cite{RestucciaSotomayor2016,FernandezRestucciaSotomayor2019}. Canonical formulations of space-time interchanged modified KdV systems were constructed in Ref.~\cite{FringTairaTurner2024}.
In our previous works, we applied the Dirac-Bergmann algorithm to nonlinear Schr\"odinger equations \cite{Pazarci2023}, to a supersymmetric KdV equation \cite{PazarciSUSYKdV}, and to spatially nonlocal nonlinear equations \cite{PazarciNonlocal}. In Ref.~\cite{Pazarci2023}, we observed that the introduction of auxiliary fields affects the constraint structure. Hence, we also use an auxiliary field in the Lagrangian considered here.

The main result of this paper concerns fields periodic in the longitudinal coordinate $x$. In this case, the field $u=2\varphi_x$ has zero mean in $x$, while the action still contains the mean of the potential $\varphi$. We show that the transverse dispersion term leads to a secondary constraint on this mean. As a result, the inverse derivative is replaced by its zero mean counterpart, and the Hamiltonian of Reiman and Semenov-Tyan-Shanskii \cite{ReymanSemenov} is reproduced, up to normalization conventions. This constraint depends only on the transverse dispersion term, and it appears for a general class of KP-type equations. In addition, we classify the smooth solutions of the auxiliary-field action. We also show that the reduced bracket and Hamiltonian coincide with the classical starting point of the quantization of the KP equation on a cylinder \cite{KST2017}. Moreover, the potential Lagrangian provides an action for this formulation, in which the mean of the potential is removed by the same secondary constraint.

The kinetic term $-\frac12\varphi_t\varphi_x$ is, up to the orientation of $x$, that of the Floreanini-Jackiw chiral boson \cite{FloreaniniJackiw1987}. For decaying fields, the constraints of this model are second class \cite{CostaGirotti1988}. On a circle, however, the zero mode of the constraint is first class \cite{DevecchiGirotti1994}. With transverse dispersion, it becomes a pair of second class constraints, in analogy with the zero mode constraint of Maskawa and Yamawaki in discretized light-cone quantization \cite{MaskawaYamawaki1976,Yamawaki1998}.

The paper is organized as follows. In Section~\ref{sec:DB}, we review the Dirac-Bergmann algorithm and fix the notation. In Section~\ref{sec:KP}, we carry out the constraint analysis of the KP equation for rapidly decreasing fields. In Section~\ref{sec:periodic}, we consider fields periodic in $x$ and compare the result with that of Reiman and Semenov-Tyan-Shanskii. In Section~\ref{sec:general}, we extend the zero mode analysis to KP-type equations, and Table~\ref{tab:summary} summarizes the three settings. In Section~\ref{sec:disc}, we discuss the relation to the quantization of the KP equation and open problems, and we conclude in Section~\ref{sec:conc}. In the appendices, we give the Faddeev-Jackiw cross-check, two remarks on the rapidly decreasing setting and on the auxiliary field, and the details of the limit of infinite period.

%=====================================================================
\section{The Dirac-Bergmann algorithm}
\label{sec:DB}
%=====================================================================

In this section, we briefly summarize the Dirac-Bergmann procedure for field theories \cite{Dirac1964,HenneauxTeitelboim1992}, mainly to fix our notation. Let $\mathcal{L}(q_a,\pd_t q_a,\pd_i q_a,\dots)$ be a Lagrangian density depending on the fields $q_a$, $a=1,\dots,N$, and their derivatives. The canonical momenta are
\begin{equation}
\pi_a=\frac{\pd \mathcal{L}}{\pd(\pd_t q_a)} .
\end{equation}
If the Hessian matrix
\begin{equation}
K_{ab}=\frac{\pd^2\mathcal{L}}{\pd(\pd_t q_a)\,\pd(\pd_t q_b)}
\end{equation}
has rank $R<N$, the momenta are not independent functions of the velocities and there are $N-R$ relations
\begin{equation}
c_\alpha(q,\pi)\approx 0 ,
\end{equation}
called primary constraints. Here, $\approx$ denotes weak equality, i.e., equality on the constraint surface. The canonical part of the Hamiltonian density is obtained by the Legendre transformation,
\begin{equation}
\mathcal{H}_L=\pi_a\,\pd_t q_a-\mathcal{L},
\end{equation}
and the total Hamiltonian density is
\begin{equation}
\mathcal{H}=\mathcal{H}_L+\mathcal{H}_c ,\qquad \mathcal{H}_c=\lambda_\alpha c_\alpha ,
\end{equation}
where $\mathcal{H}_c$ is the contribution of the constraints and $\lambda_\alpha$ are Lagrange multipliers. The time evolution of any functional $F$ is given by $\pd_t F=\{F,H\}$, where $H=\int d\vx\,\mathcal{H}$, with the equal-time Poisson brackets
\begin{equation}
\{q_a(\vx),\pi_b(\vx')\}=\delta_{ab}\,\delta(\vx-\vx') .
\label{eq:PB}
\end{equation}

The constraints must be preserved in time, $\pd_t c_\alpha=\{c_\alpha,H\}\approx0$. These consistency conditions either determine some of the Lagrange multipliers, or are identically satisfied, or lead to new relations among the phase space variables, called secondary constraints. The secondary constraints are also added to $\mathcal{H}_c$ with their own Lagrange multipliers. The consistency conditions are then imposed on the full set of constraints. The procedure is repeated until no new constraints arise. The complete set of constraints $\{c_A\}$ is then divided into first class constraints, whose Poisson brackets with all constraints vanish weakly, and second class constraints. For second class constraints, the matrix $C_{AB}(\vx,\vx')=\{c_A(\vx),c_B(\vx')\}$ is invertible. Then, the constraints can be imposed strongly, once the Poisson bracket is replaced by the Dirac bracket
\begin{equation}
\{F,G\}_D=\{F,G\}-\int d\mathbf{z}\,d\mathbf{w}\,\{F,c_A(\mathbf{z})\}\,C^{-1}_{AB}(\mathbf{z},\mathbf{w})\,\{c_B(\mathbf{w}),G\} .
\label{eq:Dirac}
\end{equation}
For systems with only second class constraints, the equations of motion can then be written as $\pd_t F\approx\{F,H_L\}_D$, with $H_L=\int d\vx\,\mathcal{H}_L$ evaluated on the constraint surface.

In Sections~\ref{sec:DB} and \ref{sec:KP}, and in Appendices~A, B and C unless stated otherwise, we assume that the fields and their derivatives vanish sufficiently fast at spatial infinity. Hence, boundary terms can be dropped. Integrations by parts and inverse derivatives are understood on admissible fields. These are rapidly decreasing fields for which all boundary terms vanish and the inverse derivatives of the fields that appear below decay at infinity. For the potential, this requires $\int\varphi\,dx=0$. Admissibility is a condition on the fields at a fixed time. Indeed, the space of rapidly decreasing potentials is not preserved by the evolution, except for the zero solution (see \ref{subsec:domain}). Sections~\ref{sec:periodic} and \ref{sec:general} deal with fields periodic in $x$. Throughout the paper, the symbol $\dpx$ denotes the inverse derivative
\begin{equation}
\big(\dpx f\big)(x,y)=\frac12\left(\int_{-\infty}^{x}-\int_{x}^{\infty}\right)f(x',y)\,dx' ,
\label{eq:inverse}
\end{equation}
which satisfies $\pd_x\dpx f=f$ and is antisymmetric,
\begin{equation}
\int f\,\dpx g\,dx=-\int g\,\dpx f\,dx .
\label{eq:antisym}
\end{equation}

%=====================================================================
\section{KP equation for rapidly decreasing fields}
\label{sec:KP}
%=====================================================================

In this section, we carry out the constraint analysis of the KP equation for rapidly decreasing fields. First, we introduce the Lagrangian and determine the constraints and the Lagrange multipliers. Then, we obtain Hamilton's equations, the Dirac brackets and the reduced Hamiltonian. We discuss the decay of the potential in \ref{subsec:domain}, and we compare the results with the analysis without the auxiliary field in \ref{subsec:noaux}. The Faddeev-Jackiw cross-check is given in \ref{app:FJ}.

\subsection{The Lagrangian}

We consider the KP equation in the form
\begin{equation}
u_t=u_{xxx}-6uu_x+3\sigma^2\dpx u_{yy},
\label{eq:KP}
\end{equation}
where $\sigma^2=\pm1$ distinguishes the two versions of the equation. After $x\to-x$ and $u\to-u$, Eq.~\eqref{eq:KP} takes the standard form
\begin{equation}
(u_t+6uu_x+u_{xxx})_x+3\sigma^2u_{yy}=0 ,
\label{eq:KPstd}
\end{equation}
so $\sigma^2=-1$ corresponds to KP-I and $\sigma^2=1$ to KP-II. Introducing the potential $v$ through $u=2v_x$, Eq.~\eqref{eq:KP} becomes the local equation
\begin{equation}
v_{xt}-v_{4x}+12v_xv_{xx}-3\sigma^2v_{yy}=0 ,
\label{eq:KPpot}
\end{equation}
which can be derived from a variational principle. A Lagrangian of this form, with a different normalization, appears in Ref.~\cite{Marikhin2015}. Setting $v=\varphi$, we consider the Lagrangian density
\begin{equation}
\mathcal{L}=-\frac12\varphi_t\varphi_x-\frac12\varphi_{xx}^2-4\psi\varphi_x^2+2\psi^2\varphi_x+\frac32\sigma^2\varphi_y^2 ,
\label{eq:LKP}
\end{equation}
where $\psi$ is an auxiliary field. The Euler-Lagrange equations of \eqref{eq:LKP} read
\begin{align}
\varphi_{xt}-\varphi_{4x}+8\left(\psi\varphi_x\right)_x-4\psi\psi_x-3\sigma^2\varphi_{yy}&=0 ,\label{eq:ELphi}\\
4\varphi_x\left(\psi-\varphi_x\right)&=0 .\label{eq:ELpsi}
\end{align}
Wherever $\varphi_x\neq0$, the second equation gives $\psi=\varphi_x$. The points where $\varphi_x$ vanishes are discussed in the next subsection. Substituting $\psi=\varphi_x$ into \eqref{eq:ELphi}, we obtain the potential KP equation \eqref{eq:KPpot}. In other words, on the branch $\psi=\varphi_x$ selected below, the auxiliary field coincides with the first derivative of the potential. The cubic term of the KP Lagrangian, $-2\varphi_x^3$, is then recovered from the terms of \eqref{eq:LKP} that contain $\psi$.

The Lagrangian \eqref{eq:LKP} is already of first order in time, so $\psi$ is not needed to remove higher time derivatives. As in Ref.~\cite{Pazarci2023}, we keep it to see how $\psi=\varphi_x$ emerges as a constraint and whether it changes the Hamiltonian structure (\ref{subsec:noaux}). Different first-order realizations of second-order degenerate Lagrangians were compared in a similar way in Ref.~\cite{CagatayEsenGumral2019}.

\subsection{Constraint analysis}

The configuration space consists of the two fields $\varphi$ and $\psi$. Since the Lagrangian \eqref{eq:LKP} is linear in $\varphi_t$ and independent of $\psi_t$, the Hessian matrix is the zero matrix. The difference between its dimension and its rank is two, so we have two primary constraints. The canonical momenta are
\begin{equation}
\pi_\varphi=\frac{\pd\mathcal{L}}{\pd\varphi_t}=-\frac12\varphi_x,\qquad
\pi_\psi=\frac{\pd\mathcal{L}}{\pd\psi_t}=0 ,
\label{eq:momenta}
\end{equation}
which yield the primary constraints
\begin{align}
c_1&=\pi_\varphi+\frac12\varphi_x\approx0 ,\label{eq:c1}\\
c_2&=\pi_\psi\approx0 .\label{eq:c2}
\end{align}
The canonical part of the Hamiltonian density is obtained by the Legendre transformation,
\begin{equation}
\mathcal{H}_L=\pi_\varphi\varphi_t+\pi_\psi\psi_t-\mathcal{L}
=\frac12\varphi_{xx}^2+4\psi\varphi_x^2-2\psi^2\varphi_x-\frac32\sigma^2\varphi_y^2 ,
\label{eq:HL}
\end{equation}
and the total Hamiltonian density is
\begin{equation}
\mathcal{H}=\mathcal{H}_L+\mathcal{H}_c
=\frac12\varphi_{xx}^2+4\psi\varphi_x^2-2\psi^2\varphi_x-\frac32\sigma^2\varphi_y^2
+\lambda_1\left(\pi_\varphi+\frac12\varphi_x\right)+\lambda_2\pi_\psi ,
\label{eq:HT}
\end{equation}
where the last two terms form the contribution $\mathcal{H}_c$ of the constraints.

The fundamental Poisson brackets are those of Eq.~\eqref{eq:PB} for the fields $\varphi$ and $\psi$, where $\vx=(x,y)$ and $\D=\delta(x-x')\delta(y-y')$. For later use, we record the functional derivatives of $H_L$,
\begin{align}
\frac{\delta H_L}{\delta\varphi}&=\varphi_{4x}-8\left(\psi\varphi_x\right)_x+4\psi\psi_x+3\sigma^2\varphi_{yy} ,\label{eq:dHdphi}\\
\frac{\delta H_L}{\delta\psi}&=4\varphi_x\left(\varphi_x-\psi\right) .\label{eq:dHdpsi}
\end{align}

The conservation of $c_2$ under time evolution gives
\begin{equation}
\pd_t c_2=\{c_2,H\}=-\frac{\delta H_L}{\delta\psi}=4\varphi_x\left(\psi-\varphi_x\right)\approx0 .
\end{equation}
This condition does not contain any Lagrange multiplier, so it is a secondary constraint. 
This constraint defines the union of two branches, $\varphi_x=0$ and $\psi=\varphi_x$, which meet where $\varphi_x=\psi=0$. It is irregular in the sense of Ref.~\cite{MiskovicZanelli2003}. Points where $\varphi_x$ vanishes are not exceptional. For instance, for fields periodic in $x$, the condition $\int u\,dx=0$ discussed in Section~\ref{sec:periodic} forces $u=2\varphi_x$ to vanish somewhere in every period. 

Multilinear constraints of this type split into regular branches \cite{MiskovicZanelli2003}. Replacing the product by one of its factors is therefore a choice of branch, not an equivalent reformulation. In fact, for fields periodic in $x$, the configurations $\varphi=0$, $\psi=g(y,t)$, with $g$ arbitrary, solve both Euler-Lagrange equations \eqref{eq:ELphi} and \eqref{eq:ELpsi}. However, they satisfy $\psi=\varphi_x$ only for $g=0$. The auxiliary-field action therefore has solutions outside the branch that reproduces the KP equation. We select this branch, which we call the KP branch, i.e., we take the regular secondary constraint
\begin{equation}
\ct=\psi-\varphi_x\approx0 .
\label{eq:ct}
\end{equation}

The remaining question is whether $\psi$ can differ from $\varphi_x$ in a region where $\varphi_x$ vanishes identically.
For smooth solutions, the field equations settle this question. Let $r=\psi-\varphi_x$. By \eqref{eq:ELpsi}, the product of $r$ and $\varphi_x$ vanishes, and Eq.~\eqref{eq:ELphi} becomes the potential KP equation with the source $2\pd_x(r^2)$. Where $r$ is nonzero, the potential does not depend on $x$. The left-hand side of \eqref{eq:KPpot} then reduces to its transverse term. Hence, the second $x$ derivative of $r^2$ vanishes there. It also vanishes where $r$ is identically zero. These two types of region fill the domain up to their boundaries, so by continuity $r^2$ is linear in $x$ everywhere.
For fields decaying in $x$, this forces $r=0$, so every smooth solution lies on the branch $\psi=\varphi_x$. For fields periodic in $x$, $r^2$ is constant in $x$. Since $r$ is continuous, $r$ itself is independent of $x$, the source vanishes, and the potential still solves the potential KP equation. However, $r$ can be nonzero wherever the potential is constant over a whole period. Thus, solutions outside the KP branch exist only for periodic fields, and they differ from it by an $x$-independent mode of the auxiliary field. All results below refer to the branch $\psi=\varphi_x$. With this choice, the contribution of the constraints is modified as
\begin{equation}
\mathcal{H}_c\equiv\lambda_1c_1+\lambda_2c_2+\lt\ct ,
\label{eq:Hcmod}
\end{equation}
and the Lagrange multipliers $\lambda_1$, $\lambda_2$ and $\lt$ are to be determined through the consistency conditions of all three constraints.
The nonvanishing Poisson brackets among the constraints are
\begin{align}
\{c_1(\vx),c_1(\vx')\}&=\pd_x\D ,\nonumber\\
\{c_1(\vx),\ct(\vx')\}&=-\pd_x\D ,\label{eq:algebra}\\
\{c_2(\vx),\ct(\vx')\}&=-\D .\nonumber
\end{align}
Using \eqref{eq:algebra}, the consistency conditions read
\begin{align}
\pd_t c_1&=\{c_1,H\}=-\frac{\delta H_L}{\delta\varphi}+\pd_x\lambda_1-\pd_x\lt\approx0 ,\label{eq:cons1}\\
\pd_t c_2&=\{c_2,H\}=-\frac{\delta H_L}{\delta\psi}-\lt\approx0 ,\label{eq:cons2}\\
\pd_t \ct&=\{\ct,H\}=\lambda_2-\pd_x\lambda_1\approx0 .\label{eq:cons3}
\end{align}
Eq.~\eqref{eq:cons2} determines $\lt$, Eq.~\eqref{eq:cons1} then determines $\lambda_1$, and Eq.~\eqref{eq:cons3} determines $\lambda_2$. Using \eqref{eq:dHdphi} and \eqref{eq:dHdpsi}, we obtain
\begin{align}
\lambda_1&=\varphi_{3x}-4\varphi_x^2-4\psi\varphi_x+2\psi^2+3\sigma^2\dpx\varphi_{yy} ,\label{eq:lambda1}\\
\lambda_2&=\varphi_{4x}-8\varphi_x\varphi_{xx}-4\varphi_x\psi_x-4\psi\varphi_{xx}+4\psi\psi_x+3\sigma^2\varphi_{yy} ,\label{eq:lambda2}\\
\lt&=4\varphi_x\left(\psi-\varphi_x\right) .\label{eq:lambdat}
\end{align}
The integration leading to \eqref{eq:lambda1} produces an arbitrary function $a(y,t)$, which is fixed by the boundary conditions. With the antisymmetric inverse \eqref{eq:inverse}, decay of the velocity $\varphi_t=\lambda_1$ at both ends requires $a=0$, together with the condition discussed in \ref{subsec:domain}. The origin of this freedom is discussed in Section~\ref{sec:periodic}. 
Since all multipliers are determined, no further constraints arise and the algorithm terminates. We note that the multiplier $\lt$ is proportional to $\ct$ itself and therefore vanishes on the constraint surface. The nonlocal operator $\dpx$ in $\lambda_1$ comes from the transverse dispersion term and is absent in the KdV case.

After a substitution of the determined Lagrange multipliers into the total Hamiltonian density, and performing integration by parts, we have
\begin{align}\nonumber
\mathcal{H}=& 3\psi^2\varphi_x-6\psi\varphi_x^2+2\varphi_x^3-\frac32\sigma^2\varphi_y^2+\frac32\sigma^2\varphi_x\dpx\varphi_{yy}\\& \label{eq:HTfull}
+\pi_\varphi\left(\varphi_{3x}-4\varphi_x^2-4\psi\varphi_x+2\psi^2+3\sigma^2\dpx\varphi_{yy}\right)\\&
+\pi_\psi\left(\varphi_{4x}-8\varphi_x\varphi_{xx}-4\varphi_x\psi_x-4\psi\varphi_{xx}+4\psi\psi_x+3\sigma^2\varphi_{yy}\right) .
\nonumber
\end{align}
Hamilton's equations follow from the total Hamiltonian as
\begin{align}
&\varphi_t+4\varphi_x^2-\varphi_{3x}+4\psi\varphi_x-2\psi^2-3\sigma^2\dpx\varphi_{yy}=0 ,\label{eq:H1}\\
&\psi_t+8\varphi_x\varphi_{xx}-\varphi_{4x}+4\varphi_x\psi_x+4\psi\varphi_{xx}-4\psi\psi_x-3\sigma^2\varphi_{yy}=0 ,\label{eq:H2}
\\
&\varphi_{xt}-\varphi_{4x}+12\varphi_x\varphi_{xx}-3\sigma^2\varphi_{yy}=0 ,
\label{eq:H3}
\\
& \varphi_x\left(\psi-\varphi_x\right)=0,
\label{eq:H4}
\end{align}
where we substituted the canonical momenta and used $\psi=\varphi_x$ in the last two equations. These equations are equivalent to the Euler-Lagrange equations \eqref{eq:ELphi} and \eqref{eq:ELpsi}. Hence, they give the KP equation \eqref{eq:KP} in terms of $u=2\varphi_x=2\psi$.

\subsection{Dirac brackets and the reduced Hamiltonian}

Since the constraint matrix $C_{AB}$ of Section~\ref{sec:DB}, with $c_A=(c_1,c_2,\ct)$, is invertible, all three constraints are second class. From \eqref{eq:algebra}, we have
\begin{equation}
C(\vx,\vx')=
\begin{pmatrix}
\pd_x & 0 & -\pd_x\\
0 & 0 & -1\\
-\pd_x & 1 & 0
\end{pmatrix}\D ,
\end{equation}
and its inverse, in the sense of integral kernels, is
\begin{equation}
C^{-1}(\vx,\vx')=
\begin{pmatrix}
\dpx & -1 & 0\\
1 & -\pd_x & 1\\
0 & -1 & 0
\end{pmatrix}\D .
\end{equation}
Here, $\dpx$ is the antisymmetric inverse \eqref{eq:inverse}. Since it maps a rapidly decreasing function to a function with nonzero limits in general, this inverse is understood on admissible fields, as stated in Section~\ref{sec:DB}. Using the definition \eqref{eq:Dirac}, the Dirac brackets of the fields are
\begin{align}
\{\varphi(\vx),\varphi(\vx')\}_D&=\dpx\D ,\nonumber\\
\{\varphi(\vx),\psi(\vx')\}_D&=-\D ,\label{eq:DiracKP}\\
\{\psi(\vx),\psi(\vx')\}_D&=-\pd_x\D .\nonumber
\end{align}
These brackets are consistent with the strong imposition of the constraints. For instance, the last two brackets follow from the first one by setting $\psi=\varphi_x$. The momenta are eliminated through \eqref{eq:momenta}, and $\psi$ through $\psi=\varphi_x$. Hence, the reduced phase space is described by $\varphi$ alone, with the bracket in the first line of \eqref{eq:DiracKP}.

If one substitutes the canonical momenta and uses the constraint $\psi=\varphi_x$, the total Hamiltonian density \eqref{eq:HTfull} becomes, up to total derivatives,
\begin{align}\nonumber
\mathcal{H}&=\frac12\varphi_{xx}^2+2\varphi_x^3-\frac32\sigma^2\varphi_y^2
\\&=
\frac12\psi_x^2+2\psi^3-\frac32\sigma^2\left(\dpx\psi_y\right)^2 .
\label{eq:Hred}
\end{align}
One can easily check that the Dirac brackets reproduce the equation of motion. Indeed,
\begin{equation}
\varphi_t=\{\varphi,H\}_D=\dpx\frac{\delta H}{\delta\varphi}
=\varphi_{3x}-6\varphi_x^2+3\sigma^2\dpx\varphi_{yy} .
\end{equation}

Finally, in terms of the KP field $u=2\varphi_x=2\psi$, the Hamiltonian \eqref{eq:Hred} reads
\begin{equation}
H=\int d\vx\left(\frac18u_x^2+\frac14u^3-\frac38\sigma^2\left(\dpx u_y\right)^2\right) ,
\label{eq:HKPu}
\end{equation}
and the Dirac bracket \eqref{eq:DiracKP} gives
\begin{equation}
\{u(\vx),u(\vx')\}_D=-4\,\pd_x\D .
\label{eq:uuD}
\end{equation}
The KP equation is then written in Hamiltonian form as
\begin{equation}
u_t=-4\,\pd_x\frac{\delta H}{\delta u}
=-4\,\pd_x\left(-\frac14u_{xx}+\frac34u^2-\frac34\sigma^2\pd_x^{-2}u_{yy}\right)
=u_{xxx}-6uu_x+3\sigma^2\dpx u_{yy} ,
\end{equation}
which is in accordance with the well-known Hamiltonian structure of the KP equation \cite{NMPZ1984,ReymanSemenov}, up to the normalization of the Poisson operator. When the transverse term is dropped, Eq.~\eqref{eq:HKPu} reduces to the Hamiltonian of the KdV equation with the Gardner-Zakharov-Faddeev Poisson operator \cite{Gardner1971,ZakharovFaddeev1971}.

%=====================================================================
\section{Periodic fields and the zero mode constraint}
\label{sec:periodic}
%=====================================================================

\subsection{Boundary conditions and notation}

As emphasized by Reiman and Semenov-Tyan-Shanskii \cite{ReymanSemenov}, a consistent Hamiltonian treatment of the KP equation is possible for functions that are periodic, or almost periodic, in $x$. For rapidly decreasing functions, it meets serious obstacles. In this section, we repeat the constraint analysis of Section~\ref{sec:KP} for fields periodic in $x$ with period $\ell$. We show that the zero modes of $\pd_x$, which were disposed of by boundary conditions in Section~\ref{sec:KP}, now lead to an additional secondary constraint. In the $y$ direction, we assume, for definiteness, that the fields vanish at infinity, and we comment on the periodic case below. This is the setting of the semi-periodic initial value problem for the KP-II equation, which has been studied by the inverse spectral transform under the zero-mass constraint $\int u\,dx=0$ \cite{Kalamvokas2023}. Fields periodic in the transverse variable have been considered for the KP-I equation \cite{Robert2018}.

We denote the average over a period by
\begin{equation}
\av{f}(y)=\frac1\ell\int_0^\ell f(x,y)\,dx ,
\end{equation}
and the projection onto the zero mean part by $Pf=f-\av{f}$. The operator $\pd_x$ is invertible only on functions of zero mean. We denote its inverse by $\dph$, extended to arbitrary periodic functions by $\dph f\equiv\dph Pf$, and fixed by the requirement that $\dph f$ has zero mean. It satisfies
\begin{equation}
\pd_x\dph f=\dph\pd_xf=Pf ,\qquad \int_0^\ell f\,\dph g\,dx=-\int_0^\ell g\,\dph f\,dx ,
\label{eq:dph}
\end{equation}
and its kernel is
\begin{equation}
\dph\delta_\ell(x-x')=G_\ell(x-x') ,\qquad G_\ell(s)=\frac12-\frac{s}{\ell}\quad(0<s<\ell) ,
\label{eq:Gl}
\end{equation}
where $\delta_\ell(x-x')$ is the periodic delta function and $G_\ell$ is the sawtooth function with period $\ell$.
In the limit $\ell\to\infty$, $G_\ell(s)\to\frac12\,\mathrm{sgn}(s)$, which is the kernel of the antisymmetric inverse derivative \eqref{eq:inverse} used in Section~\ref{sec:KP}. We denote $\Dx$ as any fixed antiderivative in $x$, for instance $\Dx f=\int_0^xf\,dx'$. Note that $\Dx$ is not the antisymmetric inverse \eqref{eq:inverse}. For a function of zero mean
\begin{equation}
\dph f=\Dx f-\av{\Dx f} .
\label{eq:dphdpx}
\end{equation}

We take $\varphi$ and $\psi$ periodic in $x$. Then, $u=2\varphi_x$ satisfies $\av{u}=0$, which is exactly the condition $\int_0^1u\,dx=0$ imposed in Ref.~\cite{ReymanSemenov} on the orbit associated with the KP equation. Note that $\int u\,dx$ is a Casimir of the bracket \eqref{eq:uuD}, so fixing its value amounts to the choice of a symplectic leaf.

\subsection{Constraint analysis}

The Lagrangian \eqref{eq:LKP}, the canonical momenta, the primary constraints \eqref{eq:c1} and \eqref{eq:c2}, and the canonical part \eqref{eq:HL} are unchanged. So are the brackets \eqref{eq:algebra}, with $\D$ replaced by $\delta_\ell(x-x')\delta(y-y')$. The new feature is that the operator $\pd_x$ in the first bracket of \eqref{eq:algebra} annihilates $x$-independent functions. Consequently, the integral of $c_1$ over a period,
\begin{equation}
\Pi(y)=\int_0^\ell c_1\,dx=\int_0^\ell\pi_\varphi\,dx ,
\label{eq:Pi}
\end{equation}
has vanishing brackets with $c_1$ and $\ct$. Hence, its conservation cannot be ensured by the Lagrange multiplier $\lambda_1$. Indeed, at the stage of the primary constraints we find
\begin{equation}
\pd_t\Pi=\int_0^\ell\left(-\frac{\delta H_L}{\delta\varphi}+\pd_x\lambda_1\right)dx=-3\sigma^2\pd_y^2\int_0^\ell\varphi\,dx ,
\end{equation}
since all other terms in \eqref{eq:dHdphi} are total $x$-derivatives. This leads, in addition to $\ct$, to the secondary constraint
\begin{equation}
\chi(y)=\pd_y^2\int_0^\ell\varphi\,dx\approx0 .
\label{eq:chi}
\end{equation}

Note that $\chi\approx0$ is the condition under which the consistency condition of $c_1$ can be solved in periodic functions. It plays a role analogous to the solvability condition that fixes the constants in the normal form lemma of Ref.~\cite{ReymanSemenov}. In the rapidly decreasing case of Section~\ref{sec:KP}, this condition does not arise from the algorithm, but it is necessary for rapidly decreasing potentials to remain rapidly decreasing (\ref{subsec:domain}).
Moreover, $\chi$ is not the zero-mass constraint $\av{u}=0$. That constraint holds automatically, since $u=2\varphi_x$ and $\varphi$ is periodic. Instead, $\chi$ constrains the mean of the potential, which does not appear in $u$.

The constraint $\chi$ is analogous to the zero mode constraint found by Maskawa and Yamawaki in discretized light-cone quantization \cite{MaskawaYamawaki1976,Yamawaki1998}. There, the kinetic term is also linear in a single spatial derivative, and the zero mode of the field does not have a kinetic term. The zero mode is determined by the spatial average of its Euler-Lagrange equation, which appears as a secondary second class constraint. In interacting light-front theories, the zero mode constraint generally depends on the other modes nonlinearly. However here, $\chi$ is linear because all nonlinear terms of $\delta H_L/\delta\varphi$ are total $x$-derivatives.

The constraint $\chi$ is a function of $y$ only, and we add it to the Hamiltonian as
\begin{equation}
H=\int d\vx\left(\mathcal{H}_L+\lambda_1c_1+\lambda_2c_2+\lt\ct\right)+\int dy\,\mu\,\chi ,
\end{equation}
with a new Lagrange multiplier $\mu(y)$. The only nonvanishing bracket of $\chi$ with the other constraints is
\begin{equation}
\{c_1(\vx),\chi(y')\}=-\pd_y^2\delta(y-y') ,
\end{equation}
and the consistency conditions become
\begin{align}
\pd_tc_1&=-\frac{\delta H_L}{\delta\varphi}+\pd_x\left(\lambda_1-\lt\right)-\mu_{yy}\approx0 ,\label{eq:pcons1}\\
\pd_tc_2&=-\frac{\delta H_L}{\delta\psi}-\lt\approx0 ,\label{eq:pcons2}\\
\pd_t\ct&=\lambda_2-\pd_x\lambda_1\approx0 ,\label{eq:pcons3}\\
\pd_t\chi&=\pd_y^2\int_0^\ell\lambda_1\,dx\approx0 .\label{eq:pcons4}
\end{align}
The average of \eqref{eq:pcons1} over a period gives $-3\sigma^2\chi/\ell-\mu_{yy}\approx0$, hence $\mu_{yy}=0$. Its solutions, $\mu=\mu_1(t)y+\mu_0(t)$, with nonzero values $\mu_1$ or $\mu_0$ are excluded by the decay conditions in $y$, which the multipliers must preserve, so $\mu=0$. Its zero mean part gives
\begin{equation}
\lambda_1=\lt+\dph\frac{\delta H_L}{\delta\varphi}+m ,
\label{eq:lambda1m}
\end{equation}
where $m(y,t)$ is the zero mode left undetermined by $\pd_x$. This zero mode is now fixed by \eqref{eq:pcons4}, which gives $\pd_y^2\left(m+\av{\lt}\right)=0$. By the same decay argument, $m=-\av{\lt}$. Together with \eqref{eq:pcons2} and \eqref{eq:pcons3}, all multipliers are determined:
\begin{align}
\lt&=4\varphi_x\left(\psi-\varphi_x\right) ,\qquad \mu=0 ,\nonumber\\
\lambda_1&=P\left[\varphi_{3x}-4\varphi_x^2-4\psi\varphi_x+2\psi^2\right]+3\sigma^2\dph\varphi_{yy} ,\label{eq:lambda1P}\\
\lambda_2&=\pd_x\lambda_1 .\nonumber
\end{align}
Note that $\mu$ vanishes, just as $\lt$ vanishes weakly. Hence, $\chi$ does not act through the term $\mu\chi$. It acts through the consistency condition \eqref{eq:pcons4}, which fixes the zero mode of $\lambda_1$, and through the restriction of the phase space. 

Comparing with \eqref{eq:lambda1}, the inverse derivative is replaced by $\dph$ and the projector $P$ removes the average of the local terms. 
This freedom was fixed by the boundary conditions in Section~\ref{sec:KP}, whereas here, it is fixed by the algorithm itself through $\chi$ and its consistency condition.

\subsection{Constraint algebra and Dirac brackets}

The constraints decouple into nonzero and zero Fourier modes in $x$. The nonzero modes of $c_1$, $c_2$ and $\ct$ form the same second class system as in Section~\ref{sec:KP}, with $\dpx$ replaced by $\dph$ in the inverse constraint matrix. The zero modes are $\Pi$, $\Pi_\psi=\int_0^\ell\pi_\psi\,dx$, $\Psi=\int_0^\ell\psi\,dx$ (the integral of $\ct$ over a period, since $\int_0^\ell\varphi_x\,dx=0$), and $\chi$, with the nonvanishing brackets
\begin{equation}
\{\Pi_\psi(y),\Psi(y')\}=-\ell\,\delta(y-y'),\qquad \{\Pi(y),\chi(y')\}=-\ell\,\pd_y^2\delta(y-y') .
\end{equation}

For fields vanishing at infinity in $y$, a function of $y$ with vanishing second derivative is zero. However, $\pd_y^2$ cannot be inverted on all such fields, since a decaying source does not in general have a decaying second antiderivative. We therefore do not use the inverse of $\pd_y^2$. Instead, we note that for such fields the constraint $\chi\approx0$ is equivalent to $\av{\varphi}\approx0$. The zero mode of $\varphi$ is then described by the two constraints $\Pi\approx0$ and $\av{\varphi}\approx0$. All constraints are second class since
\begin{equation}
 \{\Pi(y),\av{\varphi}(y')\}=-\delta(y-y').
\end{equation}

If the fields are periodic also in $y$, the constraint $\chi$ no longer fixes the mean of $\varphi$ over the whole cell. The corresponding momentum $\int\!\!\int\pi_\varphi\,dx\,dy$ then has vanishing brackets with all constraints, so it is first class. It generates the constant shifts $\varphi\to\varphi+\epsilon(t)$ that change the Lagrangian \eqref{eq:LKP} only by the total derivative and leave $u$ unchanged. Therefore, it is a gauge freedom, which can be fixed by the gauge condition $\int\!\!\int\varphi\,dx\,dy=0$. This is what remains of the first class zero mode of the Floreanini-Jackiw chiral boson on a circle \cite{DevecchiGirotti1994}. The transverse term breaks the shifts that depend on $y$, so only the constant shift survives (see Section~\ref{sec:general}).

We split the potential into its average over a period and the remaining part, $\varphi=\av{\varphi}+\tilde\varphi$, where $\tilde\varphi$ has zero mean. The average forms a second class pair with $\Pi$, so its Dirac brackets with all functionals of the fields vanish. In other words, it is not a dynamical mode. Its value is fixed by $\chi=0$ and the boundary conditions, which give $\av{\varphi}=0$ up to the gauge constant. The reduced phase space is therefore described by the zero mean field $\tilde\varphi$, with the Dirac brackets
\begin{equation}
\{\tilde\varphi(\vx),\tilde\varphi(\vx')\}_D=G_\ell(x-x')\,\delta(y-y') ,\qquad
\{u(\vx),u(\vx')\}_D=-4\pd_x\delta_\ell(x-x')\,\delta(y-y') ,
\label{eq:Dper}
\end{equation}
where $u=2\tilde\varphi_x$. The kernel $G_\ell$ is the sawtooth that inverts $\pd_x$ on functions of zero mean. In the limit $\ell\to\infty$, it tends to $\frac12\,\mathrm{sgn}(s)$, and the first bracket reduces to \eqref{eq:DiracKP}. The derivative of the sawtooth is a periodic delta function minus the constant $1/\ell$, and the second derivative removes this constant. Hence, the zero mode modifies the bracket of the potential, while $u$ keeps the local bracket of the decaying case. The Faddeev-Jackiw approach gives the same zero mode constraint and the same Dirac brackets (\ref{app:FJ}).

\subsection{Reduced Hamiltonian and the mean-normalization term}

The Hamiltonian of Ref.~\cite{ReymanSemenov} contains an additional term, which we call the mean-normalization term. Its origin can be seen directly from the canonical part of the Hamiltonian. Since the cross terms between $\av{\varphi}$ and $\tilde\varphi$ integrate to zero over a period, the transverse part of $H_L$ splits as
\begin{equation}
-\frac32\sigma^2\int_0^\ell\varphi_y^2\,dx=-\frac32\sigma^2\left(\ell\,\av{\varphi}_y^2+\int_0^\ell\tilde\varphi_y^2\,dx\right) .
\label{eq:zeromodeenergy}
\end{equation}
After integration over $y$, the first term on the right-hand side is the energy of the zero mode $\av{\varphi}$, which has no kinetic term in the Lagrangian. The constraint $\chi=\ell\,\pd_y^2\av{\varphi}\approx0$, imposed strongly by means of the Dirac bracket, sets $\av{\varphi}_y=0$ and removes this term. On the constraint surface, the Hamiltonian therefore takes the form \eqref{eq:Hred} with $\varphi=\tilde\varphi=\frac12\dph u$, so
\begin{equation}
H=\int d\vx\left(\frac18u_x^2+\frac14u^3-\frac38\sigma^2\left(\dph u_y\right)^2\right) ,
\label{eq:HKPper}
\end{equation}
and the equation of motion reads
\begin{equation}
u_t=-4\,\pd_x\frac{\delta H}{\delta u}=u_{xxx}-6uu_x+3\sigma^2\dph u_{yy} .
\label{eq:KPper}
\end{equation}

To compare with Ref.~\cite{ReymanSemenov}, we express $\dph$ through an arbitrary fixed antiderivative $\Dx$ by means of \eqref{eq:dphdpx}. This gives
\begin{align}
\dph u_{yy}&=\Dx u_{yy}-\frac1\ell\int_0^\ell dx\,\Dx u_{yy} ,\label{eq:RSterm1}\\
\int_0^\ell\left(\dph u_y\right)^2dx&=\int_0^\ell\left(\Dx u_y\right)^2dx-\frac1\ell\left(\int_0^\ell dx\,\Dx u_y\right)^2 .\label{eq:RSterm2}
\end{align}
Both right-hand sides are independent of the choice of the antiderivative, as they must be. If the potential is written as $\varphi=\frac12\Dx u$ with an arbitrary antiderivative, its average is $\av{\varphi}=\frac12\av{\Dx u}$. Then, the zero mode energy equals the last term in \eqref{eq:RSterm2}, multiplied by $\frac38\sigma^2$ and integrated over $y$. Hence, this term is precisely the subtraction of the zero mode energy required by $\chi$. 

For $\ell=1$, Eq.~\eqref{eq:RSterm1} is exactly the combination of nonlocal terms in the KP equation of Ref.~\cite{ReymanSemenov}. There, this combination does not depend on the choice of the antiderivative, and the term $-3\int dx\,\Dx u_{yy}$ is imposed by the self-consistency condition. This condition guarantees the conservation of $\int u\,dx$. Similarly, substituting \eqref{eq:RSterm2} into \eqref{eq:HKPper} reproduces the structure of their Hamiltonian $8H_4$, where
\begin{equation}
H_4=\int\left[\frac18u^3-\frac1{16}\left(\pd_xu\right)^2+\frac3{16}\left(\Dx u_y\right)^2\right]dx\,dy
-\frac3{16}\int dy\left(\int dx\,\Dx u_y\right)^2 .
\label{eq:H4RS}
\end{equation}
In particular, the coefficient of the last term is minus the coefficient of the transverse term in the first integral, as dictated by \eqref{eq:RSterm2} with $\ell=1$. For $\sigma^2=1$, our Hamiltonian \eqref{eq:HKPper} equals $-2H_4$ evaluated at $-u$. The factor and the sign reflect the different normalizations of the Poisson operator and of the field. Therefore, up to these normalization conventions, the Dirac-Bergmann analysis reproduces the structure of the Hamiltonian obtained by the Lie-algebraic approach, including the mean-normalization term.

The two derivations are completely independent. In the former, the Poisson structure follows from the second class constraints, and $\chi$ generates the mean-normalization term. In the latter, it is the Lie-Poisson bracket of pseudodifferential symbols. There, the mean-normalization term follows from the periodicity of the gauge transformation in the normal form lemma.

The relation between the two settings can be seen in the limit $\ell\to\infty$. We consider a family of periodic fields that satisfies the conditions given in \ref{app:limit}. Then, the last term in \eqref{eq:RSterm2} and the average subtracted by the projector $P$ in \eqref{eq:lambda1P} vanish. Moreover, the values of the periodic Hamiltonian \eqref{eq:HKPper} converge to those of the Hamiltonian \eqref{eq:HKPu} for decaying fields. This explains why the mean-normalization term of Ref.~\cite{ReymanSemenov} is absent in the rapidly decreasing case. It also explains why the zero-mass constraint is needed there for the transverse term of the Hamiltonian to be finite.

%=====================================================================
\section{Extension to KP-type equations}
\label{sec:general}
%=====================================================================

The analysis of Section~\ref{sec:periodic} uses only a few features of the KP Lagrangian, and it extends to a general class of equations. This is because the Euler derivative of the longitudinal energy, i.e., of the part of the energy that contains only $x$-derivatives of $\varphi$, is a total $x$-derivative. Hence, it vanishes after averaging over a period. Consider the Lagrangian density
\begin{equation}
\mathcal{L}=-\frac12\varphi_t\varphi_x-W\left(\varphi_x,\varphi_{xx},\dots,\pd_x^n\varphi\right)+\frac\kappa2\varphi_y^2 ,
\label{eq:Lgen}
\end{equation}
where $W$ is an arbitrary (for instance, polynomial) function of the $x$-derivatives of $\varphi$ and $\kappa$ is a constant. Its Euler-Lagrange equation is
\begin{equation}
\varphi_{xt}=E+\kappa\varphi_{yy} ,\qquad E=\sum_{k=1}^n\left(-\pd_x\right)^k\frac{\pd W}{\pd\left(\pd_x^k\varphi\right)} ,
\label{eq:ELgen}
\end{equation}
and in terms of $u=2\varphi_x$ it becomes
\begin{equation}
u_t=2E+\kappa\,\dpx u_{yy} ,
\label{eq:KPtype}
\end{equation}
where $E$ is a total $x$-derivative of a function of $u$ and its $x$-derivatives. The KP equation corresponds to $W=\frac12\varphi_{xx}^2+2\varphi_x^3$ and $\kappa=3\sigma^2$. The class also contains the generalized KP equations with nonlinearity $u^pu_x$, obtained from $W\propto\varphi_x^{p+2}$. It also contains KP equations with higher-order dispersion, such as the fifth-order KP equation, obtained by adding terms like $\varphi_{xxx}^2$ to $W$. These are KP-type equations with polynomial dispersion (see also Ref.~\cite{MolinetSautTzvetkov2007}).

Since $\varphi$ enters $W$ only through $\varphi_x$, the functional derivative of $\int W\,d\vx$ remains a total $x$-derivative when $W$ contains a nonlocal operator acting on $\varphi_x$. Hence, the results below also hold for such dispersion if the kinetic term is unchanged and the operations are defined on the chosen domains.

The canonical momentum gives the single primary constraint $c\approx0$, which has the same form as $c_1$ in \eqref{eq:c1}. For fields periodic in $x$, its bracket is
\begin{equation}
\{c(\vx),c(\vx')\}=\pd_x\delta_\ell(x-x')\,\delta(y-y') .
\label{eq:ccper}
\end{equation}
The canonical Hamiltonian density and its functional derivative are
\begin{equation}
    \mathcal{H}_L=W-\frac\kappa2\varphi_y^2, \quad \delta H_L/\delta\varphi=E+\kappa\varphi_{yy}.
\end{equation}
Then, the arguments of Section~\ref{sec:periodic} lead to the following result, which holds for any $W$.

\begin{enumerate}
\item The nonzero Fourier modes of $c$ in $x$ are second class, while its zero mode $\Pi$, defined in \eqref{eq:Pi}, has vanishing brackets with $c$.
\item Since $E$ is a total $x$-derivative, the conservation of $\Pi$ gives
\begin{equation}
\pd_t\Pi=-\int_0^\ell\frac{\delta H_L}{\delta\varphi}\,dx=-\kappa\,\pd_y^2\int_0^\ell\varphi\,dx .
\end{equation}
For $\kappa\neq0$, this leads to the secondary constraint \eqref{eq:chi}, which is independent of $W$.
\item For fields decaying in $y$, $\chi=0$ is equivalent to $\av{\varphi}=0$, and the pair $(\Pi,\av{\varphi})$ is second class, as in Section~\ref{sec:periodic}. The multiplier of $\chi$ vanishes, and the multiplier of $c$ is $\lambda=\dph\left(E+\kappa\varphi_{yy}\right)$, with the zero mean inverse derivative.
\item The reduced phase space is described by the zero mean field $\tilde\varphi$, with the Dirac brackets \eqref{eq:Dper}, and the reduced Hamiltonian is
\begin{equation}
H=\int d\vx\left[W\left(\frac u2,\frac{u_x}2,\dots\right)-\frac\kappa8\left(\dph u_y\right)^2\right] ,
\label{eq:Hgen}
\end{equation}
which generates \eqref{eq:KPtype}, with $\dpx$ replaced by $\dph$, through $u_t=-4\pd_x\,\delta H/\delta u$. Writing $\dph$ in terms of an arbitrary antiderivative by means of \eqref{eq:RSterm2}, the transverse term produces the mean-normalization term
\begin{equation}
\frac{\kappa}{8\ell}\int dy\left(\int_0^\ell dx\,\Dx u_y\right)^2 .
\label{eq:meannorm}
\end{equation}
This is the subtraction of the zero mode energy $-\frac{\kappa\ell}{2}\int dy\,\av{\varphi}_y^2$, required by $\chi$.
\item For $\kappa=0$, i.e., without transverse dispersion, $\pd_t\Pi=0$ identically, no secondary constraint arises, and $\Pi$ is first class. It generates the shifts $\varphi\to\varphi+\epsilon(y,t)$, which change the Lagrangian only by the total derivative. This is the first class zero mode of the Floreanini-Jackiw chiral boson on a circle \cite{DevecchiGirotti1994}. It is also the reason why the potential of KdV-type equations is defined only up to an $x$-independent function.
\end{enumerate}

These statements follow exactly as in Section~\ref{sec:periodic}, since neither the primary constraint nor the zero mode sector involves $W$. For $\kappa\neq0$, the transverse term gives the zero mode a potential energy, \eqref{eq:meannorm}, but no kinetic term. Hence, the zero mode becomes an auxiliary field fixed by a secondary constraint, as in Section~\ref{sec:periodic}. 

Consequently, for every equation of the class, a consistent Hamiltonian formulation with periodic boundary conditions in $x$ requires the zero mean projection in the equation of motion. The mean-normalization term is then fixed by $\kappa$ alone. In the limit $\ell\to\infty$, the same argument as in Section~\ref{sec:periodic} leads to the zero-mass constraint under the assumptions of \ref{app:limit}. This agrees with the fact that the zero-mass constraint is required for the Hamiltonian formulation of general KP-type equations \cite{MolinetSautTzvetkov2007}. The auxiliary field of Section~\ref{sec:KP}, on its KP branch, enters only through $x$-derivatives of $\varphi$, and as seen in Section~\ref{sec:periodic} it does not modify the zero mode sector.

\begin{table}[t]
\centering
\footnotesize
\renewcommand{\arraystretch}{1.35}
\setlength{\tabcolsep}{3pt}
\begin{tabularx}{\linewidth}{@{}
>{\raggedright\arraybackslash\hsize=1.08\hsize\linewidth=\hsize}X
>{\raggedright\arraybackslash\hsize=1.02\hsize\linewidth=\hsize}X
>{\raggedright\arraybackslash\hsize=1.24\hsize\linewidth=\hsize}X
>{\raggedright\arraybackslash\hsize=0.99\hsize\linewidth=\hsize}X
>{\raggedright\arraybackslash\hsize=0.6\hsize\linewidth=\hsize}X
>{\raggedright\arraybackslash\hsize=1.07\hsize\linewidth=\hsize}X
@{}}
\toprule
Setting & Zero mode of $c_1$ & Secondary constraint on the mean & Class & Inverse of $\pd_x$ & Mean-normalization term\\
\midrule
Decaying fields, $\kappa\neq0$ (Section~\ref{sec:KP} and \ref{subsec:domain}) & None among decaying functions & None from the algorithm. $\pd_y^2\int\varphi\,dx=0$ is necessary for decay. & All second class & $\dpx$, Eq.~\eqref{eq:inverse} & None\\
\midrule
Periodic in $x$, $\kappa\neq0$ (Sections~\ref{sec:periodic} and \ref{sec:general}) & $\Pi=\int_0^\ell\pi_\varphi\,dx$ commutes with $c_1$ & $\chi=\pd_y^2\int_0^\ell\varphi\,dx\approx0$, equivalently $\av{\varphi}=0$ & Second class, except the global mode if $y$ is periodic & $\dph$ & $\frac{\kappa}{8\ell}\int A^2\,dy$, with $A=\int_0^\ell\Dx u_y\,dx$\\
\midrule
Periodic in $x$, $\kappa=0$ (Section~\ref{sec:general}) & $\Pi$ commutes with all constraints & None & $\Pi$ first class, generates $\varphi\to\varphi+\epsilon(y,t)$ & $\dph$ & None\\
\bottomrule
\end{tabularx}
\caption{Zero mode structure of the potential Lagrangians considered in this paper. Here $\kappa$ is the transverse coupling of Eq.~\eqref{eq:Lgen}, with $\kappa=3\sigma^2$ for the KP equation. The mean-normalization term is written with an arbitrary fixed antiderivative $\Dx$, as in Eq.~\eqref{eq:RSterm2}.}
\label{tab:summary}
\end{table}

%=====================================================================
\section{Discussion}
\label{sec:disc}
%=====================================================================

The Dirac brackets obtained here provide a natural starting point for the quantization of KP-type equations. For each value of $y$, the reduced bracket of $u$ in \eqref{eq:Dper} is the Poisson bracket of a $\mathrm{U}(1)$ current. Moreover, $\chi$ removes the mean of the potential, in analogy with discretized light-cone quantization \cite{MaskawaYamawaki1976,Yamawaki1998}. A quantization of the KP equation on a cylinder was proposed in Ref.~\cite{KST2017}, in terms of nonrelativistic bosons with masses $m=1,2,\dots$, and was developed further in Ref.~\cite{PalettaTorrielli2026}. At the classical level, the two formulations coincide. For $\ell=2\pi$, the field 
\begin{equation}
    \Phi(\theta,y)=\sqrt{\frac{\pi}{2}}\,u(-\theta,y)
\end{equation} 
has the bracket
\begin{equation}
\{\Phi(\theta,y),\Phi(\theta',y')\}=2\pi\,\pd_\theta\delta_{2\pi}(\theta-\theta')\,\delta(y-y') ,
\label{eq:PhiBr}
\end{equation}
which is the Poisson bracket of Ref.~\cite{KST2017}, where the zero average is also imposed. With the measure $d\theta/2\pi$, the Hamiltonian \eqref{eq:HKPper} equals $-3\sigma^2$ times their Hamiltonian $H_2$, with
\begin{equation}
    \beta=-1/(\sigma^2\sqrt{\pi/2}), \quad \gamma=-1/(3\sigma^2), \quad \tau=-3\sigma^2t .
\end{equation}
This is in accordance with their convention that $\gamma>0$ corresponds to KP-I.

Their formulation starts from the Poisson bracket and the Hamiltonian, without an action. Such an action follows from the Lagrangian \eqref{eq:L0}, which belongs to the class \eqref{eq:Lgen}. In fact, with
\begin{equation}
    \Lambda(\theta,y)=-\sqrt{2\pi}\,\varphi(-\theta,y),
\end{equation}
so that $\Phi=\Lambda_\theta$, the action of \eqref{eq:L0} becomes
\begin{equation}
S=\int d\tau\int_0^{2\pi}\frac{d\theta}{2\pi}\int dy\left(\frac12\Lambda_\tau\Lambda_\theta-\frac{\gamma}{2}\Lambda_{\theta\theta}^2-\frac{\beta}{3}\Lambda_\theta^3-\frac12\Lambda_y^2\right) .
\label{eq:SKST}
\end{equation}
Its kinetic term gives their bracket. Since $\Lambda$ is periodic, the field $\Phi=\Lambda_\theta$ has zero average, which is the zero-average condition of Ref.~\cite{KST2017}. The mean of $\Lambda$ itself is removed by the constraint $\chi$.

The action has further advantages. Translations in $\tau$, $\theta$ and $y$ give $H_2$, $H_0$ and $H_1$ as Noether charges, up to sign. These are the conserved quantities of Ref.~\cite{KST2017}. Together with the Galilei boost $B$ of Ref.~\cite{KST2017}, they were used in Ref.~\cite{PalettaTorrielli2026} to construct the classical density of the first higher charge $H_3$. Its quantum version was then checked with an ultraviolet regularization. The Galilei transformation acts as 
\begin{equation}
    \delta\Lambda=-\nu(2\tau\Lambda_y+y\Lambda_\theta).    
\end{equation}
It changes the Lagrangian density only by a total divergence, so it is a variational symmetry of the action. The corresponding Noether charge is $B-2\tau H_1$ up to sign. Moreover, the action may serve as a starting point for a path-integral quantization on the cylinder, to be compared with the canonical quantization of Ref.~\cite{KST2017}.

The remaining questions are the operator ordering, the regularization of Ref.~\cite{PalettaTorrielli2026}, and the treatment of the global gauge mode for fields periodic also in $y$. It would also be interesting to relate this problem to the quantum KdV theory of conformal field theory \cite{BLZ1996}. A related question concerns the Bethe equations of the quantum intermediate long wave hierarchy \cite{ProchazkaWatanabe2024,Prochazka2025}.

There are several natural directions for future work. First, an invariant phase space with decaying $u$ remains to be constructed. By \ref{subsec:domain}, its potentials cannot all be rapidly decreasing, and potentials with nonzero limits as $x\to\pm\infty$ are a natural candidate. Second, the irregular constraint surface of the auxiliary-field action can be analyzed beyond smooth solutions. Third, the periodic analysis can be extended to the almost periodic case and to quasi-periodic potentials with $\int u\,dx\neq0$, which correspond to other symplectic leaves. Note that integrating Eq.~\eqref{eq:KPstd} over a period gives $\pd_y^2\int u\,dx=0$. Hence, such potentials require fields that do not decay in $y$. For fields decaying in $y$, as in the semi-periodic inverse spectral problem of Ref.~\cite{Kalamvokas2023}, the zero-mass constraint is necessary.

Another direction is to apply the method to other (2+1)-dimensional integrable systems, such as the modified KP and Davey-Stewartson equations. Our treatment of the supersymmetric KdV equation \cite{PazarciSUSYKdV} also suggests a similar analysis of supersymmetric KP equations. For BBM-type full dispersion KP equations \cite{ErbayErbayErkip2022}, the kinetic term of a potential action would contain a dispersive operator, which changes the primary constraint. The CH-KP equation \cite{CamassaFalquiSforza2026} and the Lagrangian multiforms of the KP hierarchy \cite{SleighNijhoffCaudrelier2023,FariaMartinsNijhoffRiccombeni2023} are also natural candidates. Moreover, perturbations of a periodic potential give perturbations of $u$ with zero mean, while $\chi$ fixes the mean of the potential itself. This distinction may clarify the role of mean-zero perturbations in transverse stability analyses of periodic waves \cite{HaragusLiPelinovsky2017,BhavnaKumarPandey2021}.

\section{Conclusion}
\label{sec:conc}
%=====================================================================

In this paper, we have derived the Hamiltonian formulation of the Kadomtsev-Petviashvili equation by means of the Dirac-Bergmann algorithm. The auxiliary field of the Lagrangian leads to an irregular secondary constraint. For rapidly decreasing fields all constraints are second class on the branch $\psi=\varphi_x$, and the Dirac brackets give the Poisson operator $-4\pd_x$ and the Hamiltonian \eqref{eq:HKPu}. The same results follow from the Faddeev-Jackiw method on the same branch. For decaying fields, a condition on the mean of the potential is necessary for rapidly decreasing potentials to remain rapidly decreasing. For smooth solutions, the auxiliary field equals $\varphi_x$ for decaying fields, and for periodic fields, it can differ from $\varphi_x$ only at values of $(y,t)$ for which $\varphi_x$ vanishes over the whole period.

For fields periodic in $x$, the zero modes of $\pd_x$ lead to an additional secondary constraint on the mean of the potential. This constraint replaces the inverse derivative by its zero mean counterpart. Hence, the reduced Hamiltonian is in accordance with the one of Reiman and Semenov-Tyan-Shanskii, including the mean-normalization term, up to normalization conventions. Under the assumptions of \ref{app:limit}, the values of this Hamiltonian converge to those of the Hamiltonian for decaying fields in the limit of infinite period. The same zero mode constraint appears for a general class of KP-type equations, and it depends only on the transverse dispersion term. Without transverse dispersion, the zero mode is instead first class, as for the Floreanini-Jackiw chiral boson (see Table~\ref{tab:summary}).

Finally, after the field and time rescalings of Section~\ref{sec:disc}, the reduced bracket and Hamiltonian coincide with the classical starting point of the quantization of Ref.~\cite{KST2017}. The Lagrangian \eqref{eq:L0} provides an action for it.

\section*{Acknowledgments}
The author would like to thank Ilmar Gahramanov for his guidance and support. The author acknowledges support from the Istanbul Integrability and Stringy Topics Initiative (\href{https://istringy.org}{istringy.org}), and thanks Anthropic for providing access to Claude through the Claude Team plan.

%=====================================================================
\appendix

\section{Faddeev-Jackiw analysis}
\label{app:FJ}

The results in Sections~\ref{sec:KP} and \ref{sec:periodic} can be verified independently by the symplectic method of Faddeev and Jackiw \cite{FaddeevJackiw1988}, in its iterative form \cite{BarcelosNetoWotzasek1992}. This method is equivalent to the Dirac-Bergmann algorithm when all constraints are effective \cite{GarciaPons1997,GarciaPons1998}, i.e., when their variations do not vanish on the constraint surface. 

The Lagrangian \eqref{eq:LKP} is already of first order
\begin{equation}
    \mathcal{L}=a_i\pd_t\xi_i-V.
\end{equation}
Here, the symplectic variables are $\xi=(\varphi,\psi)$, the canonical one-form is $a_\varphi=-\frac12\varphi_x$, $a_\psi=0$, and the symplectic potential is $V=\mathcal{H}_L$, given by \eqref{eq:HL}. The equations of motion read
\begin{equation}
\int d\vx'\,f_{ij}(\vx,\vx')\,\pd_t\xi_j(\vx')=\frac{\delta V}{\delta\xi_i(\vx)} ,
\label{eq:FJeom}
\end{equation}
with the symplectic two-form
\begin{equation}
f_{ij}(\vx,\vx')=\frac{\delta a_j(\vx')}{\delta\xi_i(\vx)}-\frac{\delta a_i(\vx)}{\delta\xi_j(\vx')} ,
\qquad
f^{(0)}=\begin{pmatrix}\pd_x&0\\0&0\end{pmatrix}\D .
\end{equation}
The matrix $f^{(0)}$ is singular, with a zero mode in the $\psi$ direction. Contracting the equations of motion with this zero mode gives $\delta V/\delta\psi=0$, with $\delta V/\delta\psi$ given by \eqref{eq:dHdpsi}, which is the condition obtained from the conservation of $c_2$ in the Dirac-Bergmann analysis. As discussed there, this condition is irregular. In the Faddeev-Jackiw method, such ineffective constraints must be avoided, since they may lead to a loss of dynamical information \cite{GarciaPons1998}. Therefore, we impose instead the branch $\ct\approx0$ selected in Section~\ref{sec:KP}. The agreement of the two methods found below is therefore a consistency check on this branch, not a test of the branch selection itself. 

Following Ref.~\cite{BarcelosNetoWotzasek1992}, we introduce the constraint into the kinetic part of the Lagrangian by means of a multiplier field $\eta$,
\begin{equation}
\mathcal{L}^{(1)}=-\frac12\varphi_t\varphi_x+\eta_t\,\ct-\mathcal{H}_L ,
\end{equation}
so that $\xi^{(1)}=(\varphi,\psi,\eta)$ and $a_\eta=\psi-\varphi_x$. The new symplectic two-form and its inverse are
\begin{equation}
f^{(1)}=\begin{pmatrix}\pd_x&0&\pd_x\\0&0&1\\\pd_x&-1&0\end{pmatrix}\D ,
\qquad
\left(f^{(1)}\right)^{-1}=\begin{pmatrix}\dpx&-1&0\\1&-\pd_x&-1\\0&1&0\end{pmatrix}\D .
\end{equation}
The generalized brackets $\{\xi_i,\xi_j\}=\left(f^{(1)}\right)^{-1}_{ij}$ reproduce the Dirac brackets \eqref{eq:DiracKP}. The equations of motion $\pd_t\xi_i=\left(f^{(1)}\right)^{-1}_{ij}\delta V/\delta\xi_j$ give
\begin{equation}
\varphi_t=\dpx\frac{\delta H_L}{\delta\varphi}-\frac{\delta H_L}{\delta\psi}=\lambda_1 ,\qquad
\psi_t=\pd_x\varphi_t=\lambda_2 ,\qquad
\eta_t=\frac{\delta H_L}{\delta\psi}=-\lt ,
\end{equation}
which reproduce the Lagrange multipliers \eqref{eq:lambda1}, \eqref{eq:lambda2} and \eqref{eq:lambdat}. 

The velocity of the multiplier field $\eta$ plays the role of $-\lt$ and vanishes on the constraint surface. Thus, the Faddeev-Jackiw method reproduces the Dirac-Bergmann results without canonical momenta.

For fields periodic in $x$ (Section~\ref{sec:periodic}), the entry $\pd_x\delta_\ell(x-x')$ of the symplectic two-form has an additional zero mode, namely the functions independent of $x$. Contracting the equations of motion with it gives 
\begin{equation}
    \int_0^\ell dx\, \frac{\delta V}{\delta\varphi}=3\sigma^2\chi=0,    
\end{equation}
i.e., the same secondary constraint \eqref{eq:chi}. As in the Dirac-Bergmann analysis, for fields decaying in $y$, we impose it in the equivalent form $\av{\varphi}=0$, through a further multiplier field. The corresponding block of the symplectic two-form is then a constant, and the remaining block $\pd_x$ is inverted on the nonzero modes. The inverse contains $\dph$ in place of $\dpx$, reproducing the Dirac brackets \eqref{eq:Dper}.

\section{Remark on rapidly decreasing potentials}
\label{subsec:domain}
The antisymmetric inverse \eqref{eq:inverse} maps a rapidly decreasing function to a function with nonzero limits at both ends, equal to plus and minus half its integral. On the KP branch, only the transverse term of $\lambda_1$ survives at infinity, so the velocity of a rapidly decreasing potential has the limits
\begin{equation}
\varphi_t\to\pm\frac32\sigma^2\,\pd_y^2\int\varphi\,dx\qquad (x\to\pm\infty) .
\end{equation}
Therefore, the potential stays rapidly decreasing only if this zero mode constraint holds. It has the same form as the constraint $\chi$ of Section~\ref{sec:periodic}, so the zero mode constraint is not an artifact of periodicity.
In fact, only the zero solution stays rapidly decreasing. For such a solution, integrating \eqref{eq:KPpot} over $x$ shows that the integral of $\varphi$ over $x$ vanishes. Multiplying \eqref{eq:KPpot} by $x$ and integrating over the plane, we obtain
\begin{equation}
\frac{d}{dt}\int\!\!\int\varphi\,dx\,dy=-6\int\!\!\int\varphi_x^2\,dx\,dy .
\label{eq:virial}
\end{equation}
The left-hand side vanishes, so a real solution has $\varphi_x=0$, and hence $\varphi=0$. A complete treatment with decaying $u$ therefore requires potentials that are not rapidly decreasing, for instance with nonzero limits as $x\to\pm\infty$. We leave the construction of such a phase space for future work.

\section{Remark on the auxiliary field}
\label{subsec:noaux}

Eliminating $\psi$ from \eqref{eq:LKP} on the KP branch gives the potential KP Lagrangian
\begin{equation}
\mathcal{L}_0=-\frac12\varphi_t\varphi_x-\frac12\varphi_{xx}^2-2\varphi_x^3+\frac32\sigma^2\varphi_y^2 .
\label{eq:L0}
\end{equation}
For decaying fields, this Lagrangian has a single primary constraint, of the same form as $c_1$. Its bracket with itself is invertible, so it is second class and fixes the Lagrange multiplier, and no secondary constraint appears. The Dirac bracket and the Hamiltonian are the same as in Section~\ref{sec:KP}. Thus, the auxiliary field only adds the second class pair $(c_2,\ct)$, and it does not change the reduced phase space.

For smooth solutions, the two Lagrangians differ only for fields periodic in $x$. There, the auxiliary field may differ from $\varphi_x$ by a function of $y$ and $t$ wherever $\varphi_x$ vanishes over the whole period (Section~\ref{sec:KP}). Conversely, such a function can be added to the auxiliary field of any periodic solution of the potential KP equation, and both Euler-Lagrange equations still hold. As in Ref.~\cite{Pazarci2023}, the auxiliary field generates a secondary constraint that reproduces its definition. For the fourth-order nonlinear Schr\"odinger equation, this constraint also appears there multiplied by a field, as in \eqref{eq:ELpsi}, but the resulting choice of branch was not discussed.

\section{The limit of infinite period}
\label{app:limit}

In this appendix, we discuss how the periodic results of Section~\ref{sec:periodic} approach those for decaying fields as the period grows. For periodic fields, the antisymmetric inverse \eqref{eq:inverse} cannot be used directly, since its integrals over the whole line do not converge. We use instead its analogue on the centered cell $I_\ell=[-\ell/2,\ell/2]$,
\begin{equation}
(B_\ell f)(x)=\frac12\left(\int_{-\ell/2}^{x}-\int_{x}^{\ell/2}\right)f(x')\,dx' .
\label{eq:Bl}
\end{equation}
Acting on a periodic function of zero mean, $B_\ell$ gives a periodic antiderivative, and $\dph$ differs from it only by its average. Choosing $\Dx=B_\ell$ in \eqref{eq:RSterm2}, the mean-normalization term of the Hamiltonian \eqref{eq:HKPper} becomes
\begin{equation}
\frac{3\sigma^2}{8\ell}\int A_\ell^2\,dy ,\qquad A_\ell(y)=\int_{I_\ell}B_\ell u_y\,dx .
\label{eq:Aell}
\end{equation}
On the KP branch, the average removed by the projector $P$ in \eqref{eq:lambda1P} is
\begin{equation}
-\frac{3}{2\ell}\int_{I_\ell}u^2\,dx .
\label{eq:Pavg}
\end{equation}

Consider periodic fields $u_\ell$ whose profile in one cell approaches a localized profile $u$ as $\ell\to\infty$. Both terms above are suppressed by the factor $1/\ell$. They vanish if the integral of $A_\ell^2$ grows slower than $\ell$ and the integral of $u^2$ over the cell stays finite. This alone does not guarantee that the Hamiltonian converges. For instance, the fields 
\begin{equation}
u_\ell=\ell^{-3/2}\,b(y)\cos\frac{2\pi x}{\ell} ,
\label{eq:counterfield}
\end{equation}
with $b$ smooth and localized, satisfy both conditions and vanish as $\ell\to\infty$. Their transverse energy, however, does not vanish,
\begin{equation}
\int(\dph u_{\ell,y})^2\,dx\,dy=\frac{\|b_y\|^2}{8\pi^2} .
\label{eq:counter}
\end{equation}

The Hamiltonian \eqref{eq:HKPper} has three terms: the dispersive term, the cubic term and the transverse term. It converges to \eqref{eq:HKPu} if each of them converges. To compare a periodic field with a profile on the whole line, we keep the field in one cell and set it to zero outside. It is then sufficient that
\begin{equation}
\int\left(u_{\ell,x}-u_x\right)^2d\vx\to0 ,\qquad
\int\left|u_\ell-u\right|^3d\vx\to0 ,\qquad
\int\left(\dph u_{\ell,y}-\dpx u_y\right)^2d\vx\to0 ,
\label{eq:convcond}
\end{equation}
where the derivatives and the inverse derivative are taken on the cell before the field is set to zero outside it.
The last condition also implies the zero-mass constraint of Ref.~\cite{MolinetSautTzvetkov2007}. Indeed, by \eqref{eq:inverse},
\begin{equation}
\dpx u_y\to\pm\frac12\pd_y\int u\,dx\qquad (x\to\pm\infty) .
\end{equation}
It is square integrable only if these limits vanish, i.e., if $\int u\,dx$ is independent of $y$. The decay in $y$ then gives $\int u\,dx=0$.

\section*{Declaration of generative AI and AI-assisted technologies in the manuscript preparation process}

During the preparation of this work the author used Claude (Anthropic) and ChatGPT (OpenAI). These tools were used for literature searches, the symbolic and numerical verification of calculations, an independent critical review of the manuscript, and language editing. After using these tools, the author reviewed and edited the content as needed and takes full responsibility for the content of the publication.

{\small
\bibliographystyle{elsarticle-num}
\bibliography{references}}

\end{document}